\documentclass{article}
\usepackage{LaThuileFPSpro}
\usepackage{graphicx} 
\usepackage{dcolumn}  
\usepackage{amsmath, amsthm, amssymb} 

\graphicspath{{ps/}}

\newcommand{\ra}{\rightarrow}

\newcommand{\ks}{K_S}
\newcommand{\lamst}{\Lambda(1520)}
\newcommand{\tht}{\Theta^+}

\newcommand{\thc}{\Theta_c^0}

\newcommand{\mev}{\mathrm{MeV}}
\newcommand{\gev}{\mathrm{GeV}}
\newcommand{\mevc}{\mathrm{MeV}/c}
\newcommand{\gevc}{\mathrm{GeV}/c}
\newcommand{\cm}{\mathrm{cm}}

\newcommand{\km}{K^-}
\newcommand{\kp}{K^+}

\newcommand{\mevm}{\mathrm{MeV}/c^2}
\newcommand{\gevm}{\mathrm{GeV}/c^2}
\newcommand {\cutpbeam} {$445 < p_\mathrm{beam} < 525$ MeV/c}
\newcommand {\pbeam} {$p_\mathrm{beam}$}
\newcommand {\cutpolakao} {$\Theta_p < 100^\circ$}
\newcommand {\cutpolapro} {$\Theta_K < 100^\circ$}
\newcommand {\cutrelazi} {$\Phi > 90^\circ$}

\begin{document}
\title{EXPERIMENTAL REVIEW ON PENTAQUARKS}

\author{Michael Danilov and Roman Mizuk \\ 
{\em Institute for Theoretical and Experimental Physics}\\
{\em B.Cheremushkinskaya 25}\\
{\em 117218 Moscow}\\
{\em Russia}
}

\maketitle



\baselineskip=11.6pt

\begin{abstract}
The experimental evidence for pentaquarks is reviewed and compared with 
the experiments that do not see any sign of pentaquarks. 
This paper is based on a lecture given at the 33rd ITEP Winter School
of Physics in the beginning of 2005. Results obtained since then are
summarized in the epilogue.
\end{abstract}
\newpage
\section{Introduction}

In the quark model, mesons are bound states of a quark and an
anti-quark ($\bar{q}q$), while baryons are bound states of three
quarks ($qqq$). 
The quark model successfully explains the spectrum of known hadrons.
In addition to ordinary mesons and baryons QCD predicts existence of
exotic states, such as glueballs ($gg$, $ggg$), hybrid mesons
($\bar{q}gq$) and multiquark states ($qq\bar{q}\bar{q}$,
$qqqq\bar{q}$, $qqq\bar{q}\bar{q}\bar{q}$, $qqqqqq$, \ldots). Such
exotic states were searched for experimentally since 60s, but no
unambiguous candidates were found. Experimental evidence for the
$\tht$ pentaquark in 2003 became a sensation.

\section{Theoretical aspects of the $\tht$ pentaquark}

\begin{figure}[tbh]
\centering
\begin{picture}(550,150)
\put(40,-10){\includegraphics[width=8cm]{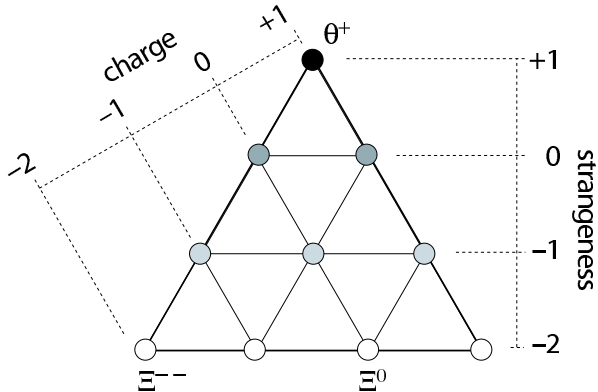}}
\end{picture}
\caption{\it The predicted anti-decuplet of pentaquark baryons.
Experimental evidence for three indicated particles has been presented.} 
\label{antidecuplet}
\end{figure}

The $\tht$ has a minimal quark content of $uudd\bar{s}$ and is the
lightest member of an untidecuplet of pentaquarks (see
Fig.~\ref{antidecuplet}) which was predicted in the chiral soliton
model\cite{chemtob}.
In 1987 Prashalovicz showed that the mass of the lightest member of
the antidecuplet is about $M=1530\,\mevm$\cite{prash_mass}. In 1997
Diakonov, Petrov and Polyakov (DPP) predicted a narrow width for this
state, $\Gamma<15\,\mev$\cite{dpp}. Narrow width is the result of the
cancellation of leading order and two subsequent orders in the $1/N_c$
expansion. The DPP motivated the experimental search for the narrow
pentaquark, and it was found in 2003 with the mass
$M=1540\,\mevm$\cite{leps,diana} and the width
$\Gamma\leq1\,\mev$\cite{cahn}. This width is extremely small for the
state, which decays strongly in a S- or P-wave and is about
$100\,\mev$ above the threshold.

The sensitivity of the $\tht$ mass prediction to the input data was
studied by Ellis, Karliner and Praszalovich\cite{ellis}. They found
that the prediction has a rather large uncertainty and the actual range
of the possible $\tht$ masses is $1430<M<1660\,\mevm$.
The ability of the chiral soliton model to explain the small width of
the $\tht$ is questioned\cite{weigel}.

The quark model predicts rather high values for pentaquark
masses\cite{strottman}.
It was attempted to explain the low mass of the $\tht$ by considering
strong correlations between quarks. According to Jaffe and
Wilczek\cite{jaffe} (Karliner and Lipkin\cite{karliner}), the $\tht$
is a combination of two diquarks and an anti-quark (diquark and
triquark). It was assumed that the correlations can considerably
decrease the mass of the pentaquarks. The low width of the $\tht$ is
explained by a low overlap of the wave functions of the pentaquark and
final $KN$ state, or by the mixing of two nearly degenerate in mass
states\cite{karliner2}.
In several recent calculation it was attempted to find precise states
of the 5-quark systems\cite{d101,d102,d103}. However, the mass values
in all the calculations were too high. Some calculations predicted
other light states, which were not experimentally observed.

Both chiral soliton model and quark model with correlated quarks
predict positive parity for the $\tht$. The QCD sum rules and Lattice
QCD, on the other hand, predict negative parity for the
$\tht$\cite{qsr,lattice}. The conclusion of the Lattice QCD is
controversial at the moment, some studies report evidence for the
$\tht$, while in most studies no low-lying narrow pentaquarks are
found. In one of the recent calculations it is shown that there exist
signals of low-lying resonances, but all of them correspond to $KN$
scattering states\cite{lattice2}.

To summarize, there is no natural explanation of the $\tht$ properties
in any of the theoretical approaches. Theory is unable to explain its
low mass and narrow width. The predictions for the parity in different
approaches are controversial.

\section{Observation of the $\tht$}

\begin{figure}[tbh]
\centering
\begin{picture}(550,140)
\put(10,-15){\includegraphics[width=11cm]{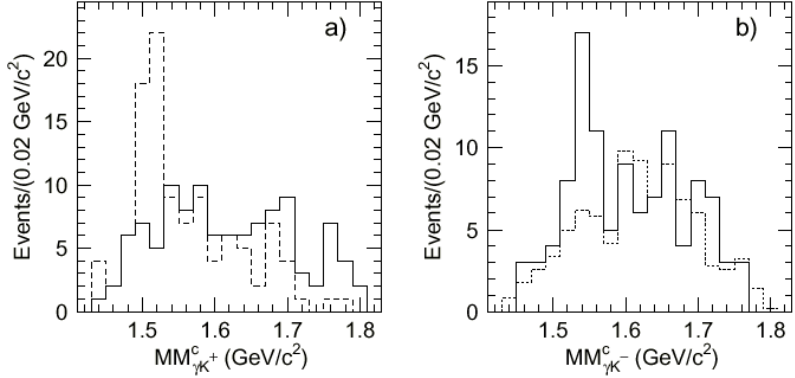}}
\end{picture}
\caption{\it Missing mass spectra for the $\gamma K^+$ (left) and
$\gamma K^-$ (right) for the reaction $\gamma C\ra
K^+K^-X$\cite{leps}. The dashed (solid) histogram shows events with
(without) additional detected proton.  The $\lamst$ signal is seen on
the left and evidence for $\tht$ is seen on the right.}
\label{LEPS}
\end{figure}

\begin{figure}[tbh]
\centering
\begin{picture}(550,120)
\put(60,-10){\includegraphics[width=7.5cm]{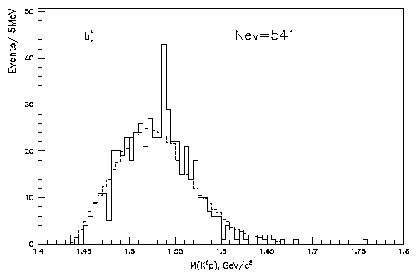}}
\end{picture}
\caption{\it Invariant mass of $pK^0$ in the reaction $K^+Xe\ra p\ks
X$\cite{diana}.  The dashed histogram is the expected background.}
\label{DIANA}
\end{figure}

Observations of a pentaquark state $\tht$ in $n\kp$\cite{leps} and
$pK^0$\cite{diana} modes created a lot of excitement.  The
corresponding invariant mass distributions obtained by the LEPS and
DIANA Collaborations are shown in Figs~\ref{LEPS}~,~\ref{DIANA}.

The minimal quark content of the $\tht$ is $uud d\overline s$.  Thus
for the first time unambiguous evidence was obtained for hadrons with
an additional quark--antiquark pair.

Analysis of the DIANA data demonstrates that the width of the $\tht$
is very small $\Gamma=0.9\pm 0.3$ MeV\cite{cahn}.  A similar small
width was obtained from the analysis of the $\kp d$ cross
section\cite{nussinov}~$^-$\cite{gibbs}.  Such a narrow width is
extremely unusual for hadronic decays and requires reassessment of our
understanding of quark dynamics. Properties of the $\tht$ were in the
excellent agreement with the theoretical predictions\cite{dpp} based
on the chiral quark soliton model. This paper motivated both
experimental searches although later on the accuracy of these
predictions was questioned\cite{ellis}.  In the quark soliton model
the $\tht$ belongs to an antidecuplet of baryons (see
Fig.~\ref{antidecuplet}).  Octet, decuplet, 27-plet, and 35-plet of
pentaquarks are also expected.

Many experiments promptly confirmed the existence of the
$\tht$\cite{clas-d}~$^-$\cite{lpi} in different processes:
photoproduction, deep inelastic scattering, hadroproduction, and
neutrino interactions. Table~\ref{positive} shows properties of the
observed peaks.

There is some spread in the mass values obtained by different
experiments. In particular masses in the $p\ks$ final state are lower
than in the $n\kp$ one. The accuracy of the mass determination is not
high in most of the experiments and therefore the disagreement is not
very serious statistically. However the DIANA and ZEUS measurements
are quite precise and contradict each other by more than 4 sigma.
Several experiments observe finite width of the $\tht$ that is much
larger than 1 MeV. However, the accuracy is again not high and within
3 sigma all width measurements are consistent with the instrumental
resolution.
\begin{center}
\begin{table}[ht]
\caption{\it Experiments with evidence for the $\tht$ baryon.}
\centering
\vskip 0.1 in
\begin{tabular}{|c|l|c|c|c|}
\hline
Reference& Group& Reaction 	& Mass 		& Width  \\
	&	&		& (MeV)		&(MeV)	  \\
\hline
\cite{leps} &
LEPS(1)	& $\gamma C \to K^+ K^- X$	& $1540\pm 10$	& $<25$	 \\
\cite{diana} &
DIANA	& $K^+ Xe \to K^0 p X$		& $1539\pm 2$	& $<9$	 \\
\cite{clas-d} &
CLAS(d)	& $\gamma d \to K^+ K^- p (n)$	& $1542\pm 5$	& $<21$	\\
\cite{saphir} &
SAPHIR	& $\gamma d \to K^+ {\overline {K^0}} (n)$	& $1540\pm 6$	& $<25$	 \\
\cite{itep} &
$\nu BC$	& $\nu A \to K^0_s p X$		& $1533\pm 5$	& $<20$	 \\
\cite{clas-p} &
CLAS	& $\gamma p\to\pi^+ K^+K^-(n)$	& $1555\pm 10$	& $<26$	 \\
\cite{hermes} &
HERMES	& $e^+ d \to K^0_S p X$		& $1526\pm 3$	& $13\pm 9$\\
\cite{zeus} &
ZEUS	& $e^+ p \to K^0_S p X$	& $1522\pm 3$	& $8\pm 4$\\
\cite{cosy} &
COSY-TOF& $p p \to K^0 p \Sigma^+$	& $1530\pm 5$	& $<18$	 \\
\cite{svd} &
SVD	& $p A \to K^0_S p X$		& $1526\pm 5$	& $<24$	  \\
\cite{leps2} &
LEPS(2) &          $\gamma d \to K^+ K^- X$ & $\sim 1530   $   & $  $   \\
\cite{itep2} &
$\nu BC2$	& $\nu A \to K^0_s p X$		& $1532\pm 2$	& $<12$	 \\
\cite{nomad}& 
NOMAD     & $\nu A \to K^0_s p X$       & $1529\pm 3$	& $<9$	 \\
\cite{jinr}& 
JINR     & $p(C_3H_8) \to K^0_s p X$       & $1545\pm 12 $	& $16\pm 4$	 \\
\cite{jinr2}& 
JINR(2)     & $CC \to K^0_s p X$       & $1532\pm 6$	& $<26$	 \\
\cite{lpi}& 
LPI     & $np \to npK^+K^-$       & $1541\pm 5$	& $<11$	 \\
\hline
\end{tabular}
\label{positive}
\end{table}
\end{center}
 
The spread in mass and width may indicate that some experiments
 observe not a signal but a statistical fluctuation. 

If the pentaquark interpretation of observed peaks is correct one
expects many other exotic (or cripto exotic) baryons belonging to the
same antidecuplet or other multiplets. Indeed several experiments
observe additional peaks in the vicinity of the $\tht$
mass\cite{itep2,jinr,lpi}.  For example three peaks with the estimated
statistical significance of 7.1, 5.0, and 4.5 $\sigma$ are seen in
neutrino interactions\cite{itep2}.
 
\begin{figure}[h!]
\begin{picture}(550,190)
\put(60,-10){\includegraphics[width=6.cm]{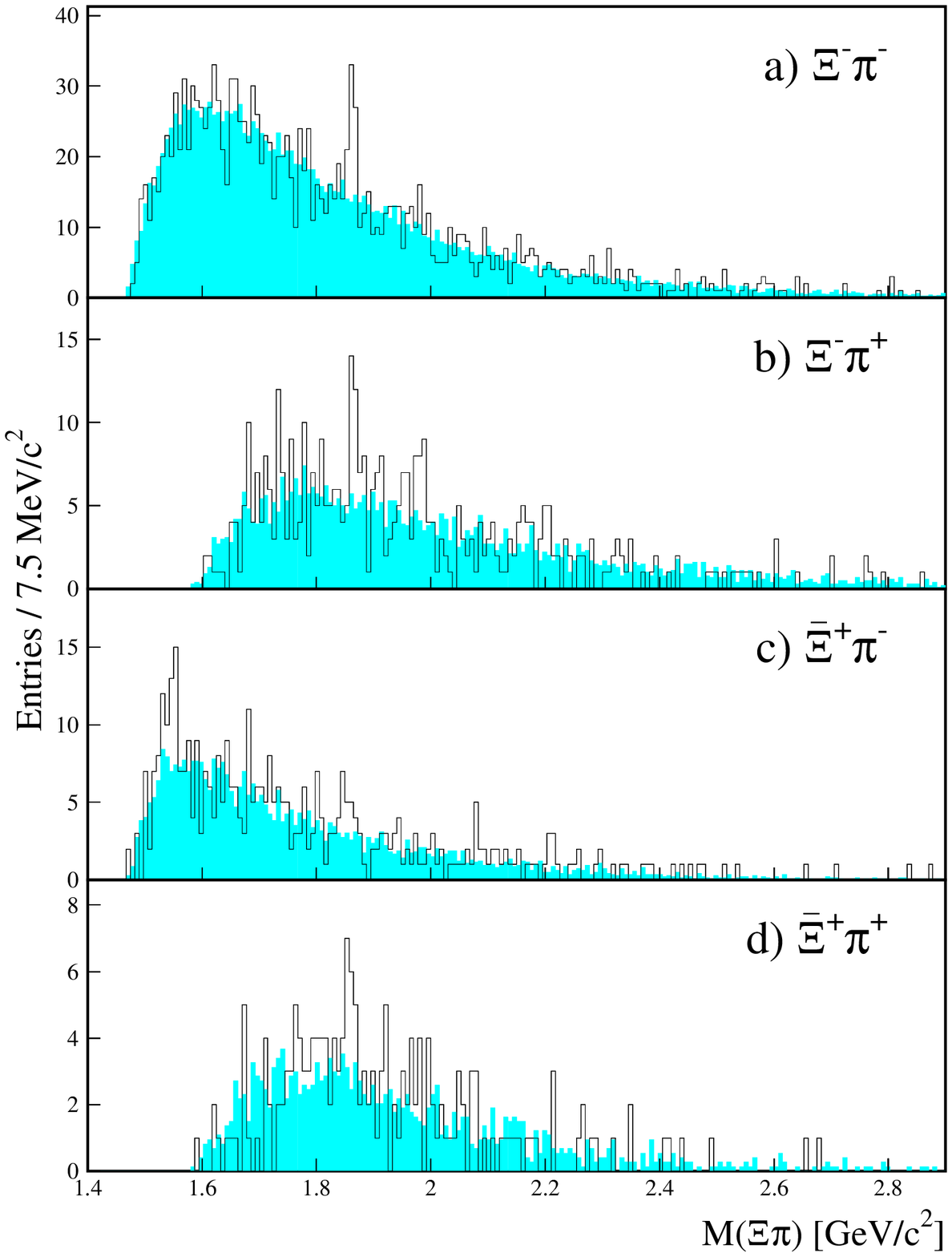}}
\end{picture}
\caption{\it Invariant mass spectra for $\Xi^-\pi^-$ (a),$\Xi^-\pi^+$
(b),$\overline{\Xi}^+\pi^-$ (c), and $\overline{\Xi}^+\pi^+$ (d) in
the NA49 experiment. The shaded histograms are the normalized
mixed-event backgrounds.}
\label{NA49}
\end{figure}

The NA49 Collaboration claims an observation of a double strange
pentaquark\cite{na49}.  Two observed narrow resonances
$\Xi^{--}_{\overline {10}}$ and $\Xi^{0}_{\overline {10}}$ (see
Fig.~\ref{NA49} )
fit naturally into the same antidecuplet as the $\tht$
(see Fig.~\ref{antidecuplet}).

\begin{figure}[h!]
\begin{picture}(550,110)
\put(40,-30){\includegraphics[width=7.0cm]{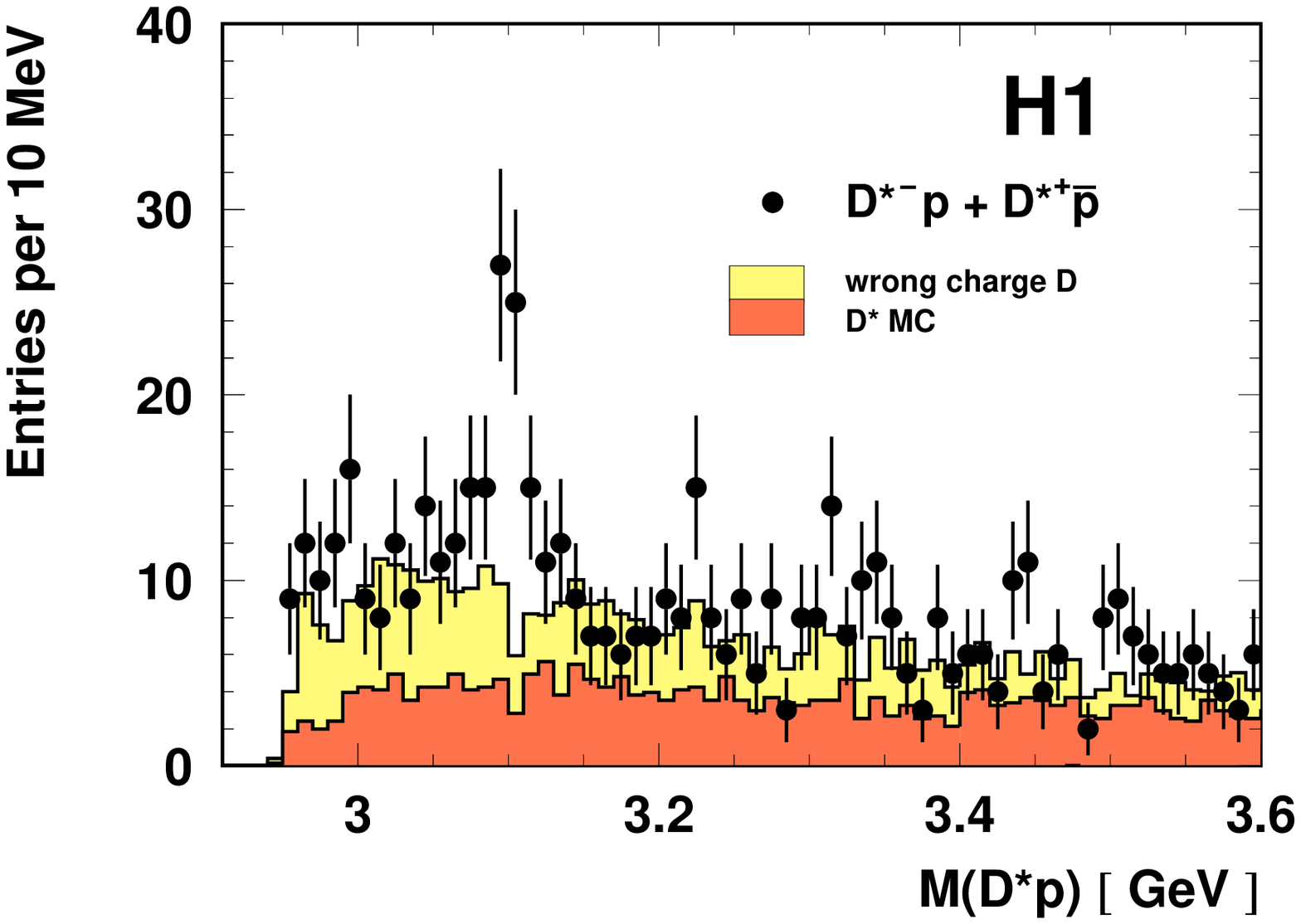}}
\end{picture}
\caption{\it Invariant mass distribution of $D^{*-}p$ and $D^{*+}\overline p$ combinations in the H1 experiment.
Two background components are shown as the shaded histograms.} 
\label{H1}
\end{figure}

An evidence for an anti-charmed pentaquark was obtained by the H1
Collaboration\cite{h1} (see Fig.~\ref{H1}).


\section{Reliability of pentaquark observations}
  
The evidence for pentaquarks was criticized by several authors (for a
review see \cite{dsierba}). They considered kinematic reflections,
ghost tracks and arbitrary selection criteria as possible explanations
for the observed peaks. The first two worries were shown to be not
important at least in some experiments (for a review
see\cite{hicks}).  The last point is especially serious since
statistical significance of the positive experiments is not high and
thus they are vulnerable to a psychological bias. This problem is
illustrated by the JINR analysis \cite{jinr} in which authors without
any reason discard the momentum range where they do not see the signal.
The ZUES Collaboration does not see the signal in data with $Q^2<20$
GeV$^2$. Their justification for discarding these data is also not too
convincing. There are other examples of experiments with not well
justified cuts. On the other hand there are experiments (for example
DIANA) in which event selection criteria have high efficiency and
reasonably justified.

\begin{figure}[h!]
\centering
\begin{picture}(550,330)
\put(0,-10){\includegraphics[width=12cm]{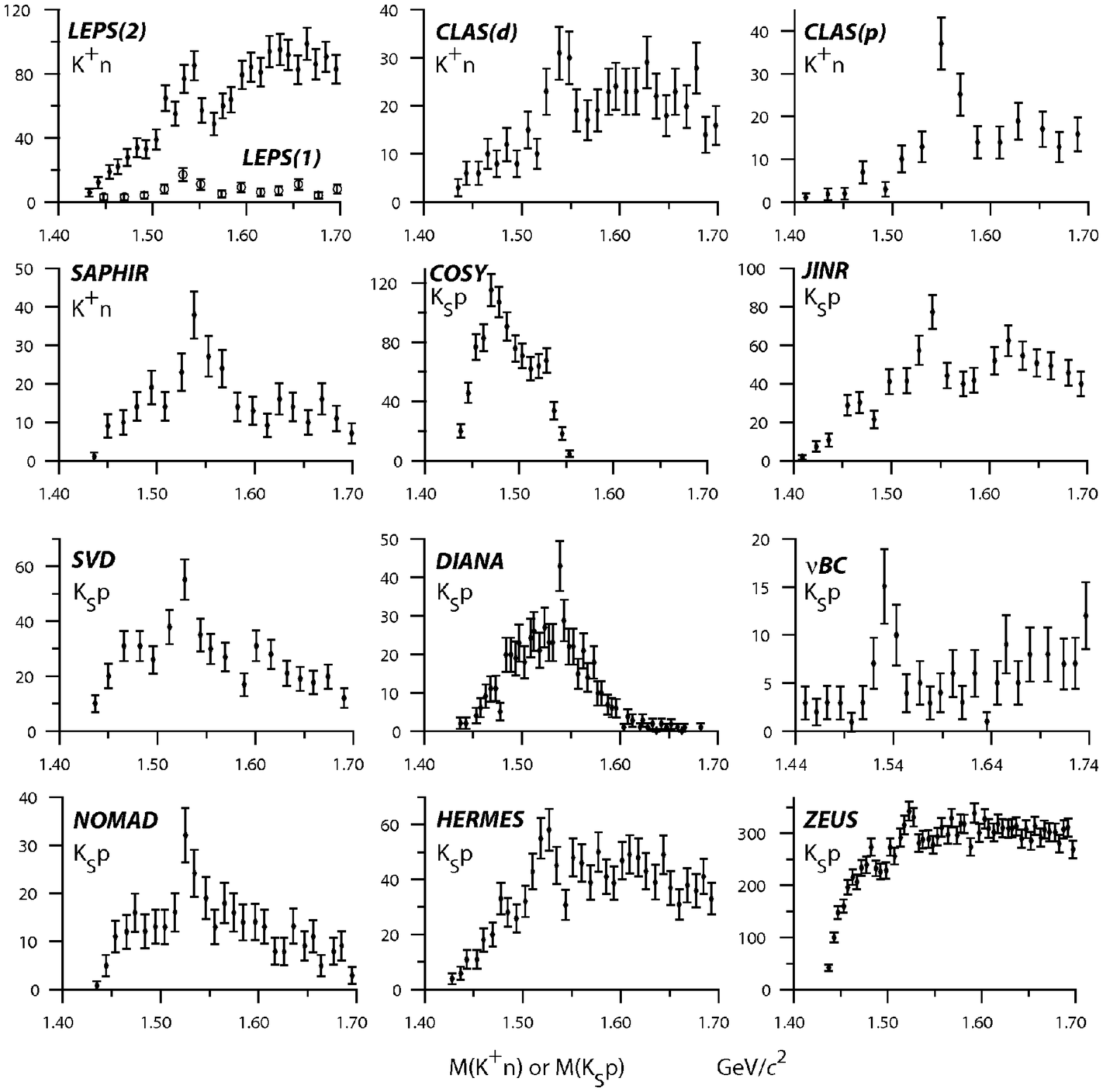}}
\end{picture}
\caption{\it Mass spectra of $nK^+$ 
 and $p\ks$ pairs in the experiments which provide evidence for the  $\tht$.} 
\label{all}
\end{figure}

The statistical significance of peaks is overestimated in all
experiments since the shape of the background is not known. This looks
obvious if one removes the fit curves and plot the data points with
error bars (see Fig.~\ref{all} taken from \cite{dsierba}).

Nevertheless the number of experiments is large and the combined
significance is high if we disregard for a moment the spread in the
peak position and width.  So one can not prove that all observed peaks
are fakes or statistical fluctuations. Only high statistics
experiments can confirm or disprove the claim for pentaquarks.

\section{Non-observation experiments}
    
Experiments which do not observe pentaquarks are shown in
Table~\ref{negative}.  Many of them are high statistics experiments
which observe by far larger number of conventional resonances than the
experiments which observe pentaquarks, and have much better mass
resolution.
\begin{table}[htb]
\caption{\it Experiments with non-observation of the $\tht$ baryon.}
\centering
\vskip 0.1 in
\begin{tabular}{|c|l|c|c|}
\hline
Reference& Group& Reaction 	& Limit \\
\hline
\cite{bes} &
BES	& $e^+e^- \to J/\Psi \to \bar{\Theta}\Theta$	& $<1.1\times 10^{-5}$ B.R.  \\
\cite{babar} &
BaBar	& $e^+e^- \to \Upsilon (4S) \to pK^0 X$		& $<1.0\times 10^{-4}$ B.R.  \\
\cite{belle} &
Belle	& $e^+e^- \to B^0\bar{B}^0 \to p\bar{p}K^0 X$	& $<2.3\times 10^{-7}$ B.R.  \\
\cite{hera-b} &
HERA-B	& $p A \to K^0_S p X$		& $<0.02 \times \Lambda^*$  \\
\cite{sphinx} &
SPHINX	& $p C \to  \Theta^+ X$	& $<0.1 \times \Lambda^*$   \\
\cite{hypercp} &
HyperCP	& $\pi ,K,p Cu \to K^0_S p X$	& $<0.3\%\ K^0p$  \\
\cite{cdf} &
CDF	& $p \bar{p} \to K^0_S p X$	& $<0.03 \times \Lambda^*$  \\
\cite{focus} &
FOCUS	& $\gamma BeO \to K^0_S p X$	& $<0.02 \times \Sigma^*$   \\
\cite{belle-h} &
Belle	& $\pi ,K,pA \to K^0_s p X$	& $<0.02 \times \Lambda^*$   \\
\cite{phenix} &
PHENIX	& $Au+Au \to K^- \bar{n} X$	& (not given)   \\
\cite{lep} &
ALEPH	& $e^+e^- \to K^0_s p X$	&  $<0.07 \times \Lambda^*$   \\
\cite{compass} &
COMPASS	& $\mu ^+A \to K^0_s p X$	&  $-$   \\
\cite{l3} &
DELPHI	& $e^+e^- \to K^0_s p X$	&  $<0.5 \times \Lambda^*$   \\
\cite{e690} &
E690	& $pp \to K^0_s p X$	&   $<0.005 \times \Lambda^*$   \\
\cite{lass} &
LASS	& $K^+p \to K^+n\pi ^+$	&  $-$   \\
\cite{l3} &
L3	& $\gamma\gamma \to K^0_s p X$	&  $<0.1 \times \Lambda$   \\
\hline
\end{tabular}
\label{negative}
\end{table}
The first significant negative result was published by the HERA-B
Collaboration\cite{hera-b}.  HERA-B does not see any evidence for the
$\tht$ but observes a clear $\lamst$ and ${\overline{\Lambda}}$(1520)
signals of about two thousand events.
HERA-B obtains an upper limit on the ratio of production cross
sections for the $\tht$ and $\lamst$ of $ R_{\Lambda^*}<2.7\%$ at the
95\% C.L. for $M_{\tht}=1530$ MeV.  In the whole range of reported
$\tht$ masses from $1522$ MeV to $1555$ MeV the limit varies up to
16\%.

The ratio of the $\tht$ and $\lamst$ production cross sections $
R_{\Lambda^*}$ is often used for the comparison of different
experiments since $\lamst$ is narrow and easily reconstructed, it has
a mass similar to the $\tht$ mass and one can draw similar diagrams
for $\lamst$ and $\tht$ production by exchanging an ${\overline{K}}$
meson into a $K$ meson.  The existence of similar diagrams
unfortunately does not prove that production mechanisms for $\tht$ and
$\lamst$ are similar. The ratio $ R_{\Lambda^*}$ is of the order of
unity in several experiments which observe the $\tht$ and less than a
few percent in many experiments which do not see $\tht$ (see
Table~\ref{negative}).

In order to resolve this discrepancy many authors assume that the
$\tht$ production drops very fast with energy and is heavily
suppressed in $e^+e^-$ annihilation. A model exists in which the
$\tht$ production cross section is strongly suppressed at high
energies in the fragmentation region\cite{titov}. It is not clear how
reliable this model is. In any case it is not applicable for the
central production for example in the HERA-B experiment where some
models predict the $\tht$ yield much higher than the experimental
limits\cite{thtprod}.
 
However, the $\tht$ production mechanism is not known and therefore it
is important to have a high statistics experiment at low energies
where most evidence for pentaquarks comes from. This goal was achieved
by the BELLE Collaboration which analyzed interactions of low momentum
particles produced in $e^+e^-$ interactions with the detector
material.  We will discuss this experiment after reviewing the
situation with the anti-charmed and doubly strange pentaquarks.
 
\section{The anti-charmed pentaquark}
       
The anti-charmed pentaquark was observed in the $p D^{*-}$ and
${\overline{p}}D^{*+}$ channels by the H1 Collaboration both in DIS
and photo production\cite{h1}.  After many experimental checks H1
concludes that the signal is real and self consistent. Still the
signal has very unusual properties. The $\thc$ measured width of
$(12\pm 3)$ MeV is consistent with the experimental resolution of
$(7\pm 2)$ MeV.  So its intrinsic width is very small although its
mass is $151$ MeV above the $p D^{*-}$ threshold and $292$ MeV above
$pD^{-}$ threshold. Its decay into $p D^{*-}$ is clearly visible
although naively one would expect much larger branching fraction for
the $p D^{-}$ channel where energy release is twice larger. Finally it
is produced with an enormous cross section. About 1.5\% of all charged
D* mesons are coming from decays of this new particle! These
properties are very surprising but we can not a priory exclude such a
possibility.
 
However, the ZEUS experiment which works at the same electron--proton
collider HERA does not see $\thc$ and gives an upper limit of 0.23\%
at the 95\% C.L. on the fraction of charged $D^*$ coming from $\thc$
decays\cite{zeus-null}.  We denote this fraction $R_{\thc/D^*}$.  For
DIS events with $Q^2>1$ GeV$^2$ the upper limit is 0.35\% at the 95\%
C.L.  This is a clear contradiction with the H1 result. We are not
aware of any convincing explanation of this discrepancy. One can try
to explain the difference using following arguments. ZEUS detects more
soft $D^*$ than H1. If one assumes that pentaquarks are produced with
high momenta only, than $D^*$ mesons from their decays should be also
energetic.  In this case soft $D^*$ that are more efficiently detected
by ZEUS should not be used in the comparison with H1.  However such an
assumption does not resolve the discrepancy since ZEUS does not see
the signal also in the kinematic range very similar to the H1 one.
 
The CDF Collaboration also does not see any sign of
$\thc$\cite{cdf}. CDF has two orders of magnitude more reconstructed
$D^*$ mesons. They reconstruct $6247\pm 1711$ $D_2^{*0}\ra
D^{*+}\pi^-$ and $3724\pm 899$ $D_1^0\ra D^{*+}\pi^-$ decays which
have the event topology very similar to $\thc$. Majority of charm
particles at HERA and Tevatron are produced in the fragmentation
process. It is impossible to reconcile the results of the two
experiments if $\thc$ is produced in the fragmentation process as
well. No other mechanism was proposed so far.  There are also upper
limits on $\thc$ production in $e^+e^-$ collisions by ALEPH\cite{lep}
and in photo production by FOCUS\cite{focus}.

We conclude that the evidence for $\thc$ is by far weaker than the
evidence against it.

 \section{Doubly strange pentaquark}

\begin{figure}[h!]
\centering
\begin{picture}(550,280)
\put(70,-10){\includegraphics[width=6cm]{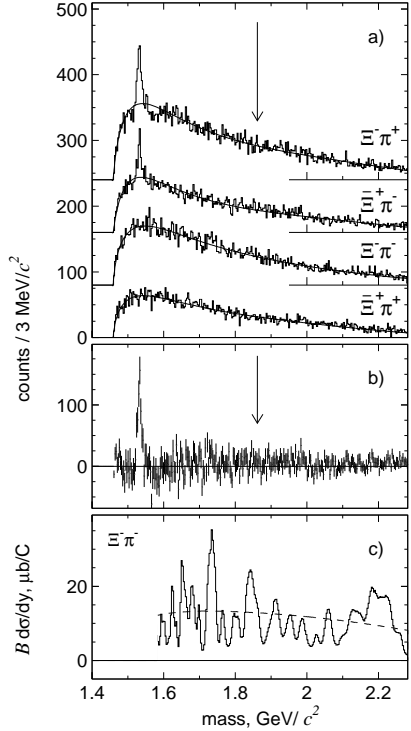}}
\end{picture}
\caption{\it The $\Xi\pi$ invariant mass spectra for $p+C$ collisions
 in the HERA-B experiment (a); sum of all four $\Xi\pi$ spectra with
 the background subtracted (b); upper limit at 95\%C.L. for mid-rapidity
 (c) .}
\label{hera-b}
\end{figure}

The NA49 claim for the observation of the doubly strange pentaquark
was not supported by several experiments which tried to find
it. HERA-B has 8 times more $\Xi^-$ hyperons and slightly better mass
resolution. There is no $\Xi(1862)$ signal in the $\Xi^-\pi^-$ or
$\Xi^-\pi^+$ mass distributions (see Fig.~\ref{hera-b}) while there is
a clear $\Xi(1530)^0$ peak with about 1000 events (including charge
conjugate combinations). HERA-B sets an upper limit of
4\%/$B(\Xi(1862)^{--}\ra\Xi^-\pi^-)$ at the 95\%C.L. on the ratio of
production cross section for $\Xi(1862)^{--}$ and $\Xi(1530)^0$. We
denote this ratio $R_{\Xi (1862)/\Xi (1530)}$.  $R_{\Xi (1862)/\Xi
(1530)}$ is about 18\%/$B(\Xi(1862)^{--}\ra\Xi^-\pi^-)$ in the NA49
experiment\cite{hera-b,na49conf}.
The center of mass energy in HERA-B is about 2 times larger than in
NA49. However the arguments about a very fast drop of the pentaquark
production cross section in the fragmentation region\cite{titov} do
not apply to the central production where the signal is observed by
NA49\cite{na49} and where it is searched for at HERA-B. The
E690 experiment has even smaller limit on the $R_{\Xi (1862)/\Xi
(1530}$ of 0.2\%/$B(\Xi(1862)^{--}\ra\Xi^-\pi^-)$ at the 95\%
C.L.\cite{e690}.  E690 studies proton--proton interactions at 800 GeV
i.e., the same process as NA49 but at the twice larger CM energy. The
WA89 experiment has about 300 times larger number of $\Xi^-$ hyperons
but does not observe $\Xi (1860)$\cite{wa89}.  However this
experiment uses a $\Sigma^-$ beam and a straightforward comparison is
not possible. The ALEPH, BaBar, CDF, COMPASS, FOCUS and ZEUS
experiments also do not see $\Xi (1862)$ in a variety of initial
processes\cite{lep,babar,cdf,compass,focus,zeus-null}.

We conclude that the evidence for $\Xi (1862)$ is by far weaker than
the evidence against it.
 
\section{The Belle experiment}

\begin{figure}[h!]
\centering
\begin{picture}(550,180)
\put(60,115){\rotatebox{90}{N/50 MeV}} 
\put(150,5){$p~(GeV)$} 
\put(55,-10){\includegraphics[width=8cm]{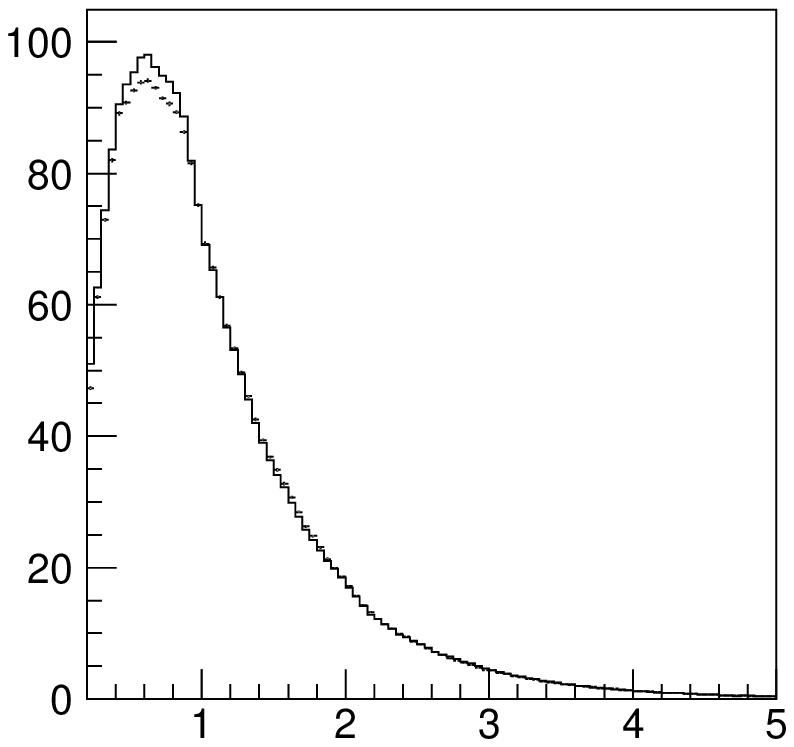}}
\end{picture}
\caption{\it Momentum spectra of $K^+$ (solid histogram) and $K^-$
(dashed histogram) in the Belle experiment.}
\label{pk}
\end{figure}

As discussed above many high statistics experiments do not see the
$\tht$ and set stringent limits on its production cross section in
different processes. It was argued, however, that the $\tht$
production can be suppressed at high energies or in specific processes
like $e^+e^-$ annihilation. Therefore Belle decided to study
interactions of low momentum particles produced in $e^+e^-$
interactions with the detector material. This allows to achieve
production conditions similar to the experiments which observe the
$\tht$ . For example the most probable kaon momentum is only 0.6 GeV
(see Fig.~\ref{pk}). The Belle kaon momentum spectrum has a large
overlap with the DIANA spectrum\cite{diana}.

\begin{figure}[h!]
\centering
\begin{picture}(550,160)
\put(3,-10){\includegraphics[width=0.95\textwidth]{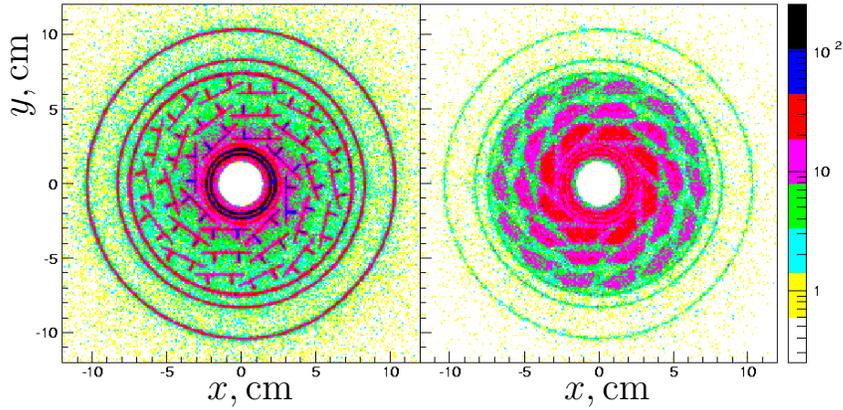}}
\put(11,110){\rotatebox{90}{\large $y,\cm$}}
\put(85,3){\large $x, \cm$} 
\put(220,3){\large $x, \cm$} 
\end{picture}
\caption{\it The $xy$ distribution of secondary $pK^-$ vertices 
for the barrel (left) and endcap (right) parts of the Belle detector.}  
\label{xy}
\end{figure}

The analysis is performed by selecting $pK^-$ and $p\ks$ secondary
vertices.  The protons and kaons are required not to originate from
the region around the run-averaged interaction point.  The proton and
kaon candidate are combined and the $pK$ vertex is fitted.  The $xy$
distribution of the secondary $pK^-$ vertices is shown in
Fig.~\ref{xy} for the barrel part (left) and for the endcap part
(right) of the detector.  The double wall beam pipe, three layers of
SVD, the SVD cover and the two support cylinders of the CDC are
clearly visible. The $xy$ distribution for secondary $p\ks$ vertices
is similar.

\begin{figure}[h!]
\centering
\begin{picture}(550,180)
\put(60,115){\rotatebox{90}{N/5 MeV}} 
\put(125,5){$M(pK^-,p\ks),$ GeV} 
\put(60,-10){\includegraphics[width=8cm]{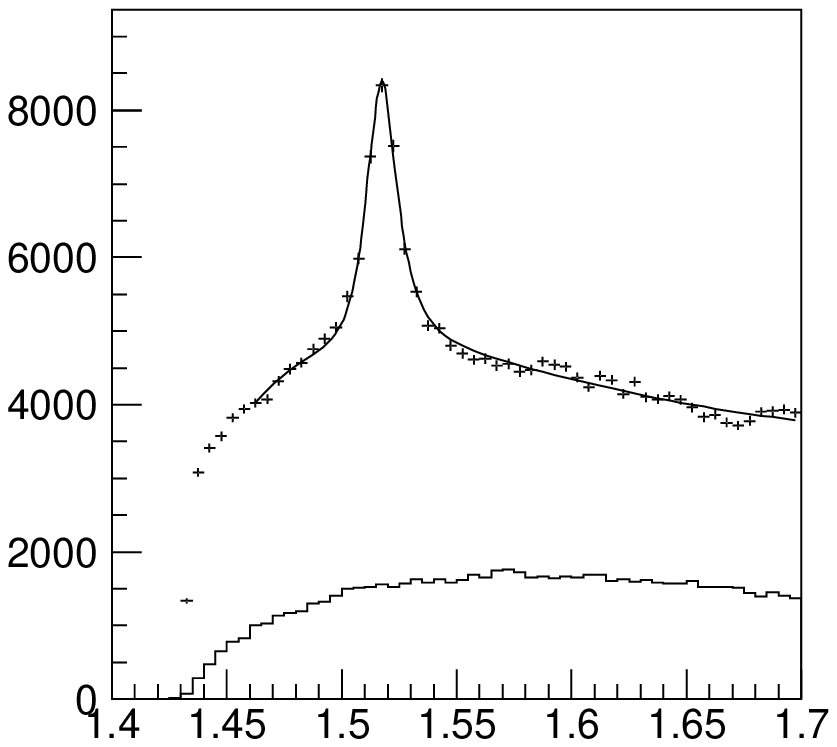}}
\end{picture}
\caption{\it Mass spectra of $pK^-$ ( points with error bars) and
$p\ks$ (histogram) secondary pairs in the Belle experiment.}
\label{m_pk}
\end{figure}

The mass spectra for $pK^-$ and $p\ks$ secondary vertices
are shown in Fig.~\ref{m_pk}. No significant structures are observed
in the $M(p\ks)$ spectrum, while in the $M(pK^-)$ spectrum a
$\lamst$ signal is clearly visible. 

\begin{figure}[h!]
\centering
\begin{picture}(550,160)
\put(60,90){\rotatebox{90}{N/200 MeV}} 
\put(145,5){p~(GeV)} 
\put(60,-10){\includegraphics[width=7cm]{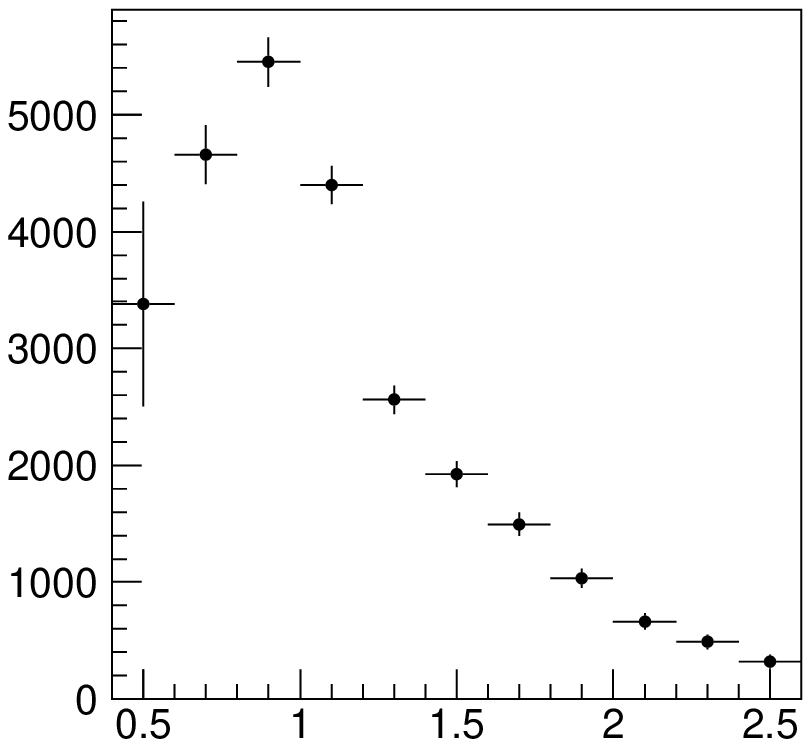}}
\end{picture}
\caption{\it $\lamst$ momentum spectrum in the Belle experiment.} 
\label{plam}
\end{figure}

The $\lamst$ yield is 15.5 thousand events. The $\lamst$ momentum
spectrum is relatively energetic (see Fig.~\ref{plam}).  $\Lambda
(1520)$ produced in a formation channel should be contained mainly in
the first bin of the histogram even in the presence of the Fermi
motion. Therefore most of $\Lambda (1520)$ are produced in the
production channel.

The upper limit for the narrow $\tht$ yield is 94 events at the 90\%
C.L. for $M_{\tht}=1540$ MeV. This leads for the upper limit of 2\% at
the 90\% C.L. on the ratio of $\tht$ and $\lamst$ production cross
sections. For other reported $\tht$ masses the limit is even smaller.

Projectiles are not reconstructed in the Belle approach. Therefore the
$\tht$ and $\lamst$ can be produced by any particle originating from
the $e^+e^-$ annihilation: $K^{\pm}$, $\pi^{\pm}$, $K^0_S$, $K^0_L$,
$p$, $\Lambda$, etc.  Belle shows that $\lamst$ are seldom accompanied
by $K^+$ mesons from the same vertex. This means that $\lamst$ are
produced mainly by particles with negative strangeness. The fraction
of energetic $\Lambda$ hyperons in $e^+e^-$ annihilation is too small
to dominate $\lamst$ production.

The Belle limit is much smaller than the results reported by many
experiments which observe the $\tht$.  For example it is two orders of
magnitude smaller than the value reported by the \mbox{HERMES}
Collaboration\cite{hermes}.  The $\tht$ and $\lamst$ are produced in
inclusive photoproduction at HERMES. Photons produce hadrons
dominantly via (virtual) pions or Kaons. Therefore the production
conditions are quite similar in the two experiments. We do not know
any physical explanation for the huge difference between the Belle and
HERMES results.

The expected number of reconstructed $\tht$ in the formation reaction
$K^+n\ra pK^0_S$ can be estimated knowing the $\tht$ width, the number
of $K^+$ mesons with appropriate momentum, amount of material and the
reconstruction efficiency. The $\tht$ width was estimated using the
DIANA data to be $0.9\pm 0.3$ MeV\cite{cahn}.  Using this value of
the $\tht$ width we estimate the number of expected $\tht$ events at
Belle to be comparable with their upper limit. If so the Belle result
disagrees with the DIANA observation. However we should wait for a
quantitative statement from the Belle Collaboration.

A comparison of the Belle upper limit on $ R_{\Lambda^*}$ with the
exclusive photoproduction experiments is not simple. However, it is
very strange to have about two orders of magnitude difference in $
R_{\Lambda^*}$ since the Belle kaon (and pion) momentum spectrum is
quite soft and comparable with the momentum spectrum of virtual kaons
(or pions) in the low energy photoproduction experiments.
     
\section{Conclusions}

The NA49 claim for the observation of $\Xi(1862)$ pentaquarks is hard
to reconcile with the results of many experiments which have up to 300
times larger statistics of usual $\Xi^-$ and $\Xi (1530)$ hyperons and
a better mass resolution. In particular E690 investigated the same
production process at about twice larger CM energy and obtained
hundred times lower limit on the ratio of $\Xi(1862)$ and $\Xi (1530)$
production cross sections.

The H1 claim for the anti-charmed pentaquark contradicts the ZEUS
study made at almost identical conditions.  CDF sets a very stringent
limit on the $\thc$ yield although they observed 178 times more $D^*$
than H1. CDF reconstructed also about 10 thousand $D_2^{*0}\ra
D^{*+}\pi^-$ and $D_1^{0}\ra D^{*+}\pi^-$ decays (including charge
conjugate states).  These decays are very similar in kinematics and
efficiency to $\thc\ra pD^{*-}$ decays (the H1 signal is observed
mainly with energetic protons for which the particle identification
does not play an important role). Three other experiments do not see
any sign of the $\thc$ in different production
processes\cite{lep,belle,focus}.  It is hard to reconcile the H1
claim with this overwhelming negative evidence.

The claims for observation of the $\tht$ in inclusive production at
medium and high energies are not supported by many high statistics
experiments which reconstruct by far larger number of ordinary
hyperons with negative strangeness. Even if one assumes that the
$\tht$ production is strongly suppressed at high energies there is
still a contradiction between several of these results with the Belle
upper limit obtained with low momentum kaons.

However, even if some claims for the $\tht$ observation are wrong it
does not mean that all observations are wrong. The DIANA and exclusive
photoproduction experiments are not in contradiction with the high
energy experiments if one assumes that the $\tht$ production drops
very fast with the energy. There is a qualitative disagreement of
these experiments with the Belle data. However here we should wait for
the quantitative analysis of the Belle data. Results of high
statistics exclusive photoproduction experiments are expected very
soon.  We hope that the situation with the pentaquark existence will
be clarified already this year.
   
\section{Epilogue}

This paper is based on a lecture given at the 33rd ITEP Winter School
of Physics in the beginning of 2005. Instead of updating the whole
text we left it unchanged in order to allow a comparison of the
situation two years ago with the present knowledge. 
The most important new results are reviewed here.

A second generation of dedicated experiments, optimized for the
pentaquark search, was undertaken at Jefferson Lab. These
photoproduction experiments cover the few-GeV beam energy range where
most of the positive evidence were reported. Each experiment collected
at least an order of magnitude more statistics than any of the
previous measurements.
No positive evidence of the $\tht$ production was reported, while two
negative results were published by the CLAS
Collaboration\cite{CLASnew1,CLASnew2}.

The CLAS Collaboration basically repeated the study of the exclusive
reaction $\gamma p\to K^0K^+n$\cite{CLASnew1} which was performed by
the SAPHIR Group\cite{saphir}. Whereas SAPHIR had reported a
$4.8\,\sigma$ significant $\tht\to n\kp$ signal, no signal was found
by CLAS.
The upper limit on the ratio of $\tht$ to $\Lambda(1520)$ production
from CLAS is more than a factor 50 lower than the value claimed by
SAPHIR Group. Thus the SAPHIR result is completely negated.

\begin{figure}[h!]
\centering
\begin{picture}(550,160)
\put(60,-10){\includegraphics[width=6.7cm]{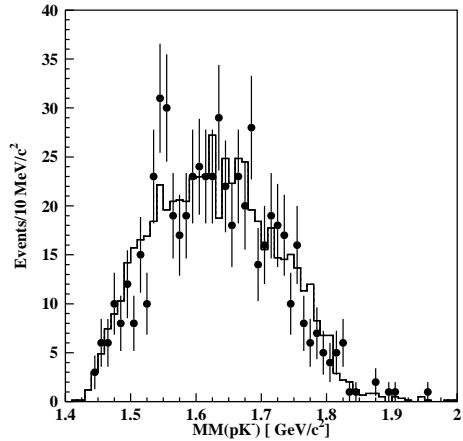}}
\end{picture}
\caption{\it Comparison of the previously published\cite{clas-d} CLAS
result (points) with the new\cite{CLASnew2} CLAS result (histogram)
normalized to get the same total number of counts.}
\label{CLASnew}
\end{figure}

The CLAS Collaboration repeated the study of the exclusive reaction
$\gamma d\to n\kp\km p$\cite{CLASnew2}. The integrated luminosity of
the new data is about a factor 30 higher than that of the previously
published CLAS paper on the same reaction, where a $4.6\,\sigma$
significant $\tht\to n\kp$ signal was claimed\cite{clas-d}. In the new
data no signal was observed. The CLAS Collaboration re-examined its
earlier work, using a background shape based on new data (see
Fig.~\ref{CLASnew}), and concluded that the background level in the
earlier sample had been underestimated and that the signal (now with
only $3\,\sigma$ significance) was probably a statistical fluctuation.

The COSY-TOF Collaboration repeated the experiment studying the
$pp\rightarrow{}pK^0\Sigma^+$ reaction with substantially improved
statistical accuracy and extended detection
capability\cite{COSYTOFnew}.
For the new measurement a slightly higher beam momentum was chosen
($3.059\,\gevc$ instead of $2.95\,\gevc$) to make the mass spectrum more
regular in the region of the expected $\tht$ signal.
No evidence for a narrow resonance in the $pK^0$ spectra was found and
the upper limit on a cross section $\sigma_{tot,X}<0.3 \mu b$
(95\%~C.L.) was set for the mass region of 1.50~GeV/c$^2$ --
1.55~GeV/c$^2$.
It was also concluded that in the previous measurement\cite{cosy} the
background level had been underestimated and that the significance of
the $\tht$ signal is much lower than claimed in the previous
publication.

In the years 2002--2003 the LEPS Collaboration collected a new data
sample of the $\gamma\,d$ and $\gamma\,p$ interactions with a factor 5
increase in statistics\cite{LEPSnew}.
Preliminary results of two analysis were reported. 
In the $K^+K^-$ detection mode the $\tht$ signal was confirmed in the
missing mass spectrum of the $K^-$ with respect to the neutron after
correcting for the Fermi-motion (the reaction is $\gamma n\to\tht
K^-\to (nK^+)K^-$). The number of the $\tht$ signal events increased
with the luminosity in the correct proportion, however, the
significance remained at the $5\,\sigma$ level.
In the second analysis the $pK^-$ final state was studied and the
$\tht$ signal was found in the missing mass spectrum of the $pK^-$
with respect to deutron requiring the $pK^-$ mass to be in the
$\lamst$ mass region (the reaction is $\gamma
d\to\tht\lamst\to\tht(pK^-)$).  The claimed significance of the peak
is $4--5\,\sigma$. The background shape is determined by two different
methods, however one has to introduce additional peak to describe the
missing mass spectrum. Thus, the background shape might be still not
well understood.
These results of LEPS are in disagreement with negative CLAS
results\cite{CLASnew1,CLASnew2}, however the two experiments have
different acceptance.
In 2006 the LEPS experiment started a new run, in which a further
increase in statistics by a factor 10 is expected. The data taking
should be finished by summer 2007. The new LEPS data might clarify the
question of agreement with CLAS.

\begin{figure}[htp]
\centering
\begin{picture}(550,160)
\put(65,110){\rotatebox{90}{$\Gamma \;(\mev)$}} 
\put(130,5){$m_{p\ks}\;(\gevm)$} 
\put(60,-10){\includegraphics[width=7cm]{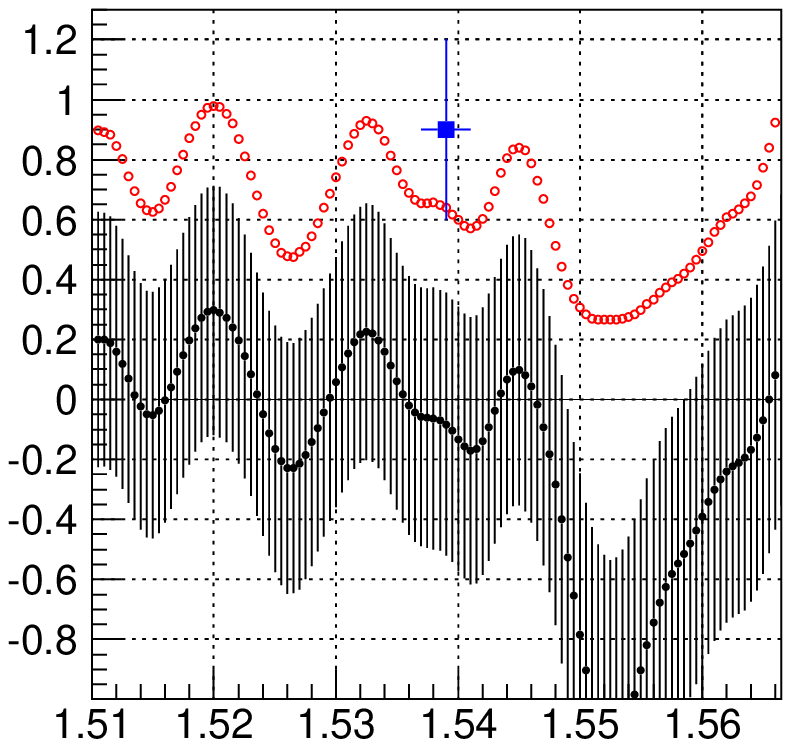}}
\end{picture}
\caption{\it Belle results for the $\tht$ yield, expressed in terms of
  the resonance width (black dots). The open dots correspond to the
  upper limit at the 90\%~C.L.. The square with error bars indicates
  the estimate made from the DIANA signal.}
\label{belle_new}
\end{figure}

The Belle Collaboration searched for the charge exchange reaction $\kp
n\to\tht\to p\ks$ using the kaon interactions in the detector
material\cite{BELLEnew}. This reaction provides a model independent
information on the width of the $\tht$ and allows to directly compare
the result with that of the DIANA Group\cite{diana}. No signal was
observed at Belle. An upper limit on the $\tht$ width
$\Gamma<0.64\,\mev$ for the $\tht$ mass $M=1.539\,\mevm$ was set, to
be compared with the estimate $\Gamma=0.9\pm0.3\,\mev$ made from the
DIANA signal\cite{cahn}. The upper limit is mass-dependent, going as
high as $1\,\mev$ for some values between 1520 and $1550\,\mevm$, as
shown in Fig.~\ref{belle_new}.
The Belle upper limit is conservative, since the $\tht$ can be
produced also in inelastic $KN$ interactions which are considered as
background in the Belle analysis.

The DIANA Group continued the investigation of the charge--exchange
reaction $\kp \mathrm{Xe}\to K^0p\,\mathrm{Xe}'$\cite{DIANAnew}.  The
statistics was almost doubled, however, the momentum distribution of
the incident $\kp$, $p_{\kp}$, in the new data was shifted to higher
values compared to the previous publication\cite{diana}. To compare
the old and new data samples, DIANA applied a requirement
$p_{\kp}<530\,\mevc$, which was automatically fulfilled for the old
data. With this requirement the total statistics increased by a factor
1.6, while the $\tht$ yield increased from 29 to $54\pm16$, thus, even
by a larger factor than the statistics.
Further reduction of the $p_{\kp}$ interval to $445<p_{\kp}<525\,\mevc$
resulted in the $\tht$ yield of $57\pm15$.

\begin{figure}[htp]
\centering
\begin{picture}(550,240)
\put(65,110){\rotatebox{90}{$\Gamma \;(\mev)$}} 
\put(130,5){$m_{p\ks}\;(\gevm)$} 
\put(0,-10){\includegraphics[width=12cm]{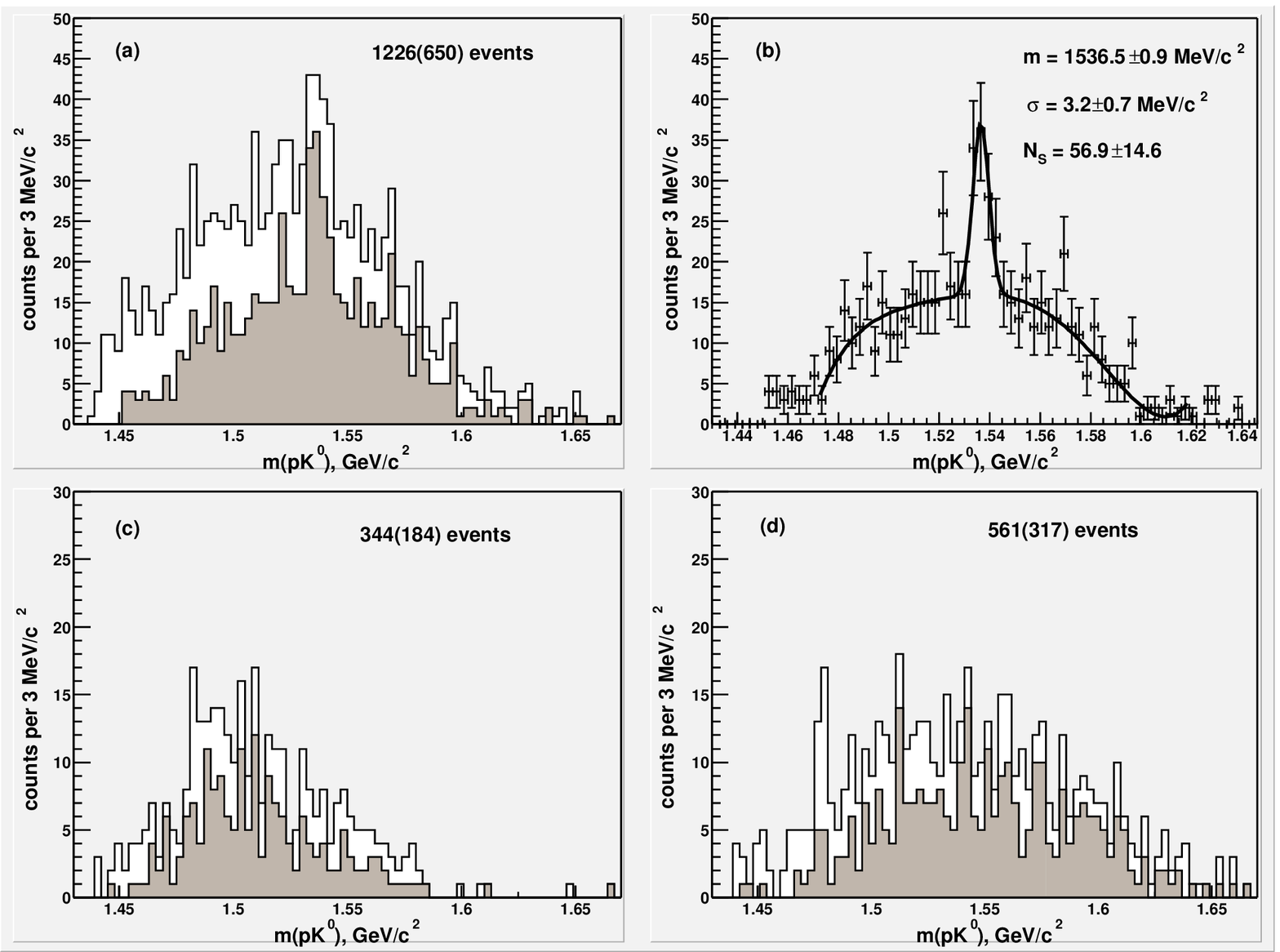}}
\end{picture}
\caption{\it Effective mass of the $K^0 p$ system at
  DIANA\cite{DIANAnew} for \cutpbeam\ (a), \pbeam\ $< 445$ MeV/c (c),
  and \pbeam\ $> 525$ MeV/c (d).  Shaded histograms are for the
  selections \cutpolakao, \cutpolapro, and \cutrelazi. Shown in (b) is
  a fit of the shaded distribution in (a) to a Gaussian on top of a
  fifth-order polynomial.}
\label{diana_new}
\end{figure}

It was found that the $\tht$ signal is concentrated in a narrow
momentum interval, $445<p_{\kp}<525\,\mevc$ (see
Fig.~\ref{diana_new}).
The fact that the $\tht$ peak is not found for $p_{\kp}>525\,\mevc$
was unexpected for the authors of this review. At
Belle\cite{BELLEnew}, the expected $\tht$ yield for
$p_{\kp}>525\,\mevc$ is roughly the same as for
$p_{\kp}<525\,\mevc$. This follows from the MC simulation, which was
verified by tagged kaon data. Based on the $p_{\kp}$ spectrum of DIANA
and Fermi-momentum spectrum of xenon nucleus, we find that the $\tht$
yield at DIANA in the $p_{\kp}>525\,\mevc$ interval should be about
30\% of the $\tht$ yield in the $445<p_{\kp}<525\,\mevc$
interval. Thus, about 17 additional events with $\tht$ should be seen
at DIANA in the $p_{\kp}>525\,\mevc$ interval. Given the background
level at DIANA, the absence of the $\tht$ signal there corresponds to
about $2.5\,\sigma$ downward fluctuation.

The new estimation of the $\tht$ width performed by DIANA is
$\Gamma=(0.36\pm0.11)\,\mev$, which is extremely small. Diakonov {\it
  et al.} conclude that with such a small width the $\tht$ can not be
produced in the photoproduction experiments and thus all the positive
evidences from such experiments can not be correct\cite{diakonov_new}.
Also the $\tht$ evidence from the analysis of the $\kp d$ cross
section\cite{nussinov}~$^-$\cite{gibbs} is negated by such a small
value of the $\tht$ width.

\begin{figure}[h!]
\centering
\begin{picture}(550,120)
\put(60,-10){\includegraphics[width=7cm]{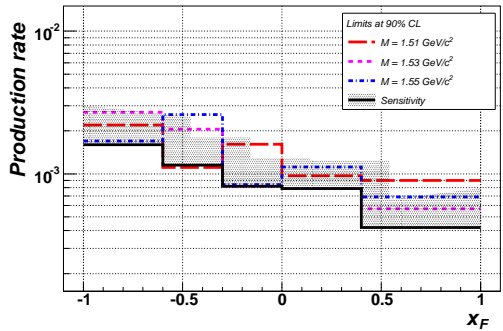}}
\end{picture}
\caption{\it Sensitivity and upper limits at 90\% C.L. for $\Theta^+$
  production rates at NOMAD\cite{NOMADnew} as a function of $x_F$,
  for $\Theta^+$ masses of 1510, 1530, 1550 MeV$/c^2$.}
\label{nomad_new}
\end{figure}

The NOMAD Collaboration searched for the $\tht$ production in the
$\nu_\mu N$ interactions\cite{NOMADnew}. The $\tht$ signal was not
observed and an upper limit on $\Theta^+$ production rate of
$2.13\cdot 10^{-3}$ per neutrino interaction (90\%~C.L.) was set.
Preliminary NOMAD results, quoting the $\tht$ signal with a
$4.3\,\sigma$ significance\cite{nomad}, suffered from an incorrect
background estimation. The results reported in\cite{nomad} were
obtained using harder proton identification requirements which yielded
an increase in the proton purity from 23\% to 51.5\% with about a
factor six loss in the statistics.
It is interesting to compare the NOMAD result\cite{NOMADnew} with the
analysis of old bubble chamber neutrino experiments which provide an
estimation of the $\Theta^+$ production rate as large as $\sim
10^{-3}$~ events per neutrino interaction\cite{itep}. As shown in
Fig.\ref{nomad_new}, for a large fraction of the $x_F$ range, except
in the region $x_F\approx -1$, such a value is excluded.

The most significant $\tht$ signal to date is from SVD-2 Collaboration,
which considerably increased the statistics and was able to confirm
its earlier observation of the $\tht$ production in the proton nucleon
interactions\cite{SVDnew}. The statistical significance of the $\tht$
signal at SVD-2 is at the level of $8\,\sigma$. The SPHINX experiment,
which operated exactly in the same environment, found null
result\cite{sphinx}. It was claimed, however, that at SVD-2 the $\tht$
is produced with very small $x_F$, while SPHINX has no acceptance in
this region. Still, it is not clear how to reconcile the SVD-2
positive result with the null result of the HERA-B
Collaboration\cite{hera-b}, which was obtained for the same reaction,
with the same acceptance in $x_F$ but with the center-of-mass energy
$40\,\gev$ instead of $12\,\gev$. The \mbox{SVD-2} yield ratio $\tht /
\lamst = 8-12\%$ is in marked disagreement with the upper limit from
HERA-B, $\tht / \lamst < 2.7\%$ (95\%~C.L.).
The CDF upper limit $\tht / \lamst < 3\%$ (90\%~C.L.)\cite{cdf} is
also applicable here, since for the central production the difference
in the nucleon-nucleon center of mass energy should not be important.

To conclude, we would like to cite a paragraph from the Particle
Data Group review on pentaquarks (edition 2006)\cite{pdg}: 
``To summarize, with the exception of SVD-2, there has not been a
high-statistics confirmation of any of the original experiments that
claimed to see the $\tht$; there have been two high-statistics repeats
from the Jefferson Lab that have clearly shown the original positive
claims in those two cases to be wrong; there have been a number of
other high-statistics experiments, none of which have found any
evidence for the $\tht$; and all attempts to confirm the two other
pentaquark states have led to negative result. The conclusion that
pentaquarks in general and the $\tht$ in particular, do not exist,
appears compelling.'' Two more negative results (COSY-TOF,
Nomad) appeared since the PDG2006 conclusion was made.

The existence of such a large number of results which were
subsequently not confirmed demonstrates the importance of the
psychological factor in the analysis, especially if the available
statistics is low.

\section{Acknowledgments}

We are grateful to A.~Kaidalov and P.~Pakhlov for the many fruitful
discussions.

\end{document}